\documentclass[a4paper]{article}

\usepackage{pxfonts}
\usepackage{bussproofs}
\usepackage{amssymb}
\usepackage{natbib}
\usepackage{url}
\usepackage{microtype}

\begin{document}

\title{A Sketch of a Proof-Theoretic Semantics for Necessity}
\author{Nils K\"urbis}
\date{}
\maketitle

\begin{center}
Published in \emph{Advances in Modal Logic 13. Booklet of Short Papers}, edited by Sara Negri, Nicola Olivetti, Rineke Verbrugge (Helsinki 2020): 37-43\\ 
\url{https://www.helsinki.fi/sites/default/files/atoms/files/finalshortpapermain.pdf}\bigskip
\end{center}

\begin{abstract}
\noindent This paper considers proof-theoretic semantics for necessity within Dummett's and Prawitz's framework. Inspired by a system of Pfenning's and Davies's, the language of intuitionist logic is extended by a higher order operator which captures a notion of validity. A notion of relative necessary is defined in terms of it, which expresses a necessary connection between the assumptions and the conclusion of a deduction. 
\end{abstract}

\section{Proof-Theoretic Semantics}
Dummett and Prawitz do not consider how the meanings of modal operators may be given by their theory of meaning for the logical constants. To investigate in outline how this may be done is the purpose of this short paper.

According to proof-theoretic semantics, the rules governing a constant define its meaning. Prior's \emph{tonk} shows that the rules cannot be arbitrary. Dummett and Prawitz impose the restriction that the introduction and elimination rules for a constant $\ast$ be \emph{in harmony}, so that $\ast E$ does not license the deduction of more consequences from $A\ast B$ than are justified by the grounds for deriving it as specified by $\ast I$. (See \citep{dummettLBM}, \citep{prawitzdummett}, \citep{prawitzdummett2}, \citep{prawitzmeaningviaproofs}, \citep{prawitzdummett3}.) A necessary condition for harmony is that deductions can be brought into normal form. A deduction is in \emph{normal form} if it contains neither maximal formulas nor maximal segments. A \emph{maximal formula} is one that is the conclusion of an $I$-rule and major premise of an $E$-rule. A \emph{maximal segment} is a sequence of formulas all except the last of which are minor premises of $\lor E$ and the last one is major premise of an $E$-rule.\footnote{I am allowing myself a certain looseness in terminology, which, however, is quite common in the literature. Strictly speaking, Dummett distinguishes intrinsic harmony, stability and total harmony. Intrinsic harmony is captured by normalisation: the elimination rules of a constant are justified relative to the introduction rules. Stability is harmony together with a suitable converse: the introduction rules are also justified relative to the elimination rules. Total harmony obtains if the constant is conservative over the rest of the language. Dummett calls the permutative reduction steps to remove maximal segments `auxiliary reduction steps'. Sometimes, as in the case of quantum disjunction, these cannot be carried out, which points to a defect in the rules for the connective from the meaning-theoretical perspective \cite[250, 289]{dummettLBM}. Dummett observes that normalisation implies that each logical constant is conservative over the rest of the language \cite[250]{dummettLBM} and conjectures that `intrinsic harmony implies total harmony in a context where stability prevails' \cite[290]{dummettLBM}. Dummett and Prawitz only count those segments as maximal that begin with the conclusion of an introduction rule. The more general notion used here is found in \citep{troelstraschwichtenberg}. It is required for philosophical reasons. For more on these issues, see \cite[Ch 2]{kurbisproofandfalsity}.} 
\newpage
Deductions in intuitionist logic \textbf{I} normalise \cite[Ch 4]{prawitznaturaldeduction}:

\begin{center} 
\bottomAlignProof
\AxiomC{$A$}
\LeftLabel{$\lor I$: \ }
\UnaryInfC{$A\lor B$}
\DisplayProof\quad
\bottomAlignProof
\AxiomC{$B$}
\UnaryInfC{$A\lor B$}
\DisplayProof\qquad\qquad
\bottomAlignProof
\AxiomC{$A\lor B$}
\AxiomC{$[A]^i$}
\noLine
\UnaryInfC{$\Pi$}
\noLine
\UnaryInfC{$C$}
\AxiomC{$[B]^j$}
\noLine
\UnaryInfC{$\Sigma$}
\noLine
\UnaryInfC{$C$}
\LeftLabel{$\lor E$: \ }
\RightLabel{$_{i ,j}$}
\TrinaryInfC{$C$}
\DisplayProof
\end{center}

\bigskip

\begin{center} 
\bottomAlignProof
\AxiomC{$[A]^i$}
\noLine
\UnaryInfC{$\Pi$}
\noLine
\UnaryInfC{$B$}
\LeftLabel{$\supset I$: \ }
\RightLabel{$_i$}
\UnaryInfC{$A\supset B$}
\DisplayProof\quad 
\bottomAlignProof
\AxiomC{$A\supset B$}
\AxiomC{$A$}
\LeftLabel{$\supset E$: \ }
\BinaryInfC{$B$}
\DisplayProof\qquad\qquad
\bottomAlignProof
\AxiomC{$\bot$}
\LeftLabel{$\bot E$: \ } 
\UnaryInfC{$C$}
\DisplayProof
\end{center}

\bigskip 

\begin{center}
\AxiomC{$A$} 
\AxiomC{$B$}
\LeftLabel{$\land I$: \ }
\BinaryInfC{$A\land  B$}
\DisplayProof\qquad\qquad
\AxiomC{$A\land  B$}
\LeftLabel{$\land E$: \ }
\UnaryInfC{$A$}
\DisplayProof\quad 
\AxiomC{$A\land  B$} 
\UnaryInfC{$B$}
\DisplayProof 
\end{center}

\noindent The constants occur only in conclusions of $I$-rules and major premises of $E$-rules. Thus the conditions for using an $I$-rule and the consequences of using an $E$-rule are given independently of the constants.

The rules of \textbf{I} exemplify Dummett's notions of \emph{full-bloodedness} and \emph{molecularity} in the theory of meaning (see \citep{dummettmeaningI}, \citep{dummettmeaningII}). A full-blooded theory of meaning characterises the knowledge of speakers in virtue of which they master a language in such a way that it exhibits how a speaker who does not yet understand an expression could acquire a grasp of it. A molecular theory of meaning does so piecemeal and specifies the meanings of the expressions of a language one group of expressions at a time. A speaker need not understand the constants of \textbf{I} in order to be informed about the conditions for the application of their $I$- and $E$-rules. To understand the grounds for deriving a formula with $\ast$ as main operator, or to understand the consequences that follow from it, a speaker only needs to grasp the meanings of some sentences, but not any sentences containing $\ast$. A speaker who does not already know the meanings of the constants of intuitionist logic could acquire a grasp of their meanings by learning the rules of inference of \textbf{I}. The rules are \emph{informative}: the grounds and consequences of a formula with $\ast$ as main operator are given without reference to $\ast$. Its meaning is specified without presupposing that $\ast$ already has meaning. 

Contrast the rules of \textbf{I} with standard rules for $\Box$ in \textbf{S4}:

\begin{center}
\AxiomC{$A$}
\LeftLabel{$\Box I$: \ } 
\UnaryInfC{$\Box A$}
\DisplayProof\qquad\qquad
\AxiomC{$\Box A$}
\LeftLabel{$\Box E$: \ } 
\UnaryInfC{$A$}
\DisplayProof
\end{center}

\noindent where in $\Box I$ all assumption on which $A$ depends have the form $\Box B$. The conditions for applying $\Box I$ are not given independently of $\Box$. Thus they presuppose that $\Box$ is meaningful. Hence they do not define its meaning. Put in terms of speakers' understanding, to be able to use $\Box I$ and to infer a formula of the form $\Box A$, a speaker already needs to know how to use formulas of the form $\Box B$ in deductions, and so the speaker already needs to know the meaning of $\Box B$. Thus a speaker could not acquire a grasp of the meaning of $\Box$ by being taught those rules. As a definition of the meaning of $\Box$, these rules are circular. The $I$-rule for $\Box$ presupposes that $\Box$ already has meaning.\footnote{Prawitz proves a normalisation theorem for intuitionist \textbf{S4} and \textbf{S5} \cite[Ch 6]{prawitznaturaldeduction}. Other such systems of intuitionist \textbf{S4} are formalised by Biermann and de Paiva \citep{biermannpaiva} and von Plato \citep{vonplatoS4}. Thus normalisation is not a sufficient condition for rules to define meaning.} 

I propose that for the rules governing $\ast$ to define its meaning, they must satisfy a \emph{Principle of Molecularity}: $\ast$ must not occur in the premises and discharged hypotheses of its $I$-rules, nor in any restrictions on their application, and $\ast$ must not occur in the minor premises and discharged hypotheses of its $E$-rules, nor in any restrictions on their application. Generalising, there should be no sequence of constants $\ast_1...\ast _n$ such that the rules for $\ast_i$ refer to $\ast_j$, $i<j$, and the rules for $\ast_n$ refer to $\ast_1$. 

A promising system of modal logic from the present perspective was formalised by Pfenning and Davies \citep{pfenningdaviesmodal}. It is based on Martin-L\"of's account of judgements. They distinguish the judgment that a proposition is \emph{true} from the judgement that a proposition is \emph{valid}. $\vdash$ is interpreted as a hypothetical judgement. Validity is defined in terms of truth and hypothetical judgements, where $\cdot$ marks an empty collection of hypotheses and $\Gamma$ are hypotheses of the form `$B \ true$': (1) If $\cdot\vdash A \ true$, then $A \ valid$; (2) If $A\ valid$, then $\Gamma\vdash A\ true$. 

Their system has axioms for the two kinds of hypotheses and rules for implication and necessity. Formulas assumed to be valid are to the left of the semi-colon, those assumed to be true to its right:\footnote{\label{footnote7}The restriction on $\Gamma$ of clause (2) of the definition is not explained further. The point may well be to avoid circularity. It is effectively lifted in the axiom $hyp^\ast$.}

\begin{center}
\bottomAlignProof
\AxiomC{}
\RightLabel{$_{hyp}$}
\UnaryInfC{$\Delta; \Gamma, A \ true, \Gamma'\vdash A \ true$}
\DisplayProof\qquad\qquad
\bottomAlignProof
\AxiomC{}
\RightLabel{$_{hyp\ast}$}
\UnaryInfC{$\Delta, B \ valid, \Delta'; \Gamma \vdash B \ true$} 
\DisplayProof\bigskip

\AxiomC{$\Delta; \Gamma, A \ true \vdash B \ true$}
\RightLabel{$_{\supset I}$}
\UnaryInfC{$\Delta; \Gamma\vdash A\supset B \ true$}
\DisplayProof\qquad
\AxiomC{$\Delta; \Gamma\vdash A \supset B \ true$}
\AxiomC{$\Delta; \Gamma\vdash A \ true$}
\RightLabel{$_{\supset E}$}
\BinaryInfC{$\Delta; \Gamma\vdash B\ true$}
\DisplayProof\bigskip
 
 \AxiomC{$\Delta; \cdot \vdash A \ true$}
 \RightLabel{$_{\Box I}$}
 \UnaryInfC{$\Delta; \Gamma \vdash \Box A \ true$}
 \DisplayProof\qquad 
 \AxiomC{$\Delta; \Gamma\vdash \Box A \ true$}
 \AxiomC{$\Delta, A \ valid; \Gamma\vdash C \ true$}
  \RightLabel{$_{\Box E}$}
 \BinaryInfC{$\Delta; \Gamma\vdash C \ true$}
 \DisplayProof
\end{center}

\noindent Call this system \textbf{JM}. It is a fragment of intuitionist \textbf{S4}. A normalisation theorem can be proved for it. Its rules satisfy the Principle of Molecularity.

\section{Modal Logic with Validity}
In this section I reformulate, extend and generalise \textbf{JM}. The reformulation is three-fold. (1) I use a system of natural deduction not in sequent calculus style. (2) As any formula in \textbf{JM} is followed by either `valid' or `true', I drop the latter and simply write `$A$'. This has a philosophical point: it accords with an account of logical inference as relating propositions, not judgements. (3) I do not treat validity as a judgement either, but as a sentential operator. The generalisation consists in the observation that validity is a relation between the assumptions and the conclusion of a deduction. The extension consists in formulating rules of inference for a higher level operator $\vdash$ for this generalised notion of validity. The rules for $\Box$ appeal to it. \textbf{HM} extends \textbf{I} by $\vdash$ and $\Box$. 

Formulas of level 0 are those of \textbf{I} extended by $\Box$. Formulas of level 1 are all formulas $B_1...B_n\vdash A$, where $B_1...B_n, A$ are formulas of level 0, for $0\leq n$. $B_1... B_n \vdash A$ can be derived if there is a deduction of $A$ from $B_1...B_n$. Applying an elimination rule for $\vdash$, this is what we should get back. We may not know how $A$ was derived from $B_1...B_n$, but as we know that there is such a deduction, the inference of $A$ from $B_1...B_n$ is valid. $\vdash$ has the following rules:  

\begin{center}
\bottomAlignProof
\AxiomC{$[B_1]^{i_1} ... [B_n]^{i_n}$}
\noLine  
\UnaryInfC{$\Pi$}
\noLine
\UnaryInfC{$A$} 
\RightLabel{$_{i_1...i_n}$}
\LeftLabel{$VI:$}
\UnaryInfC{$B_1...B_n \vdash A$}
\DisplayProof\qquad
\bottomAlignProof
\AxiomC{$\Sigma$}
\noLine
\UnaryInfC{$B_1...B_n\vdash A$}
\AxiomC{$\Xi_1$}
\noLine
\UnaryInfC{$B_1$}
\AxiomC{$...$}
\AxiomC{$\Xi_n$}
\noLine
\UnaryInfC{$B_n$}
\LeftLabel{$VE:$}
\QuaternaryInfC{$A$}
\DisplayProof
\end{center}

\noindent where $B_1...B_n$, $0\leq n$, are representatives of \emph{all} the open assumption classes of $\Pi$ in any order (as the $B$s must be of level 0, there are no open assumptions of level 1). Vacuous discharge is allowed: a representative to the left of $\vdash$ may belong to an empty assumption class of $\Pi$; this corresponds to Thinning. 

$VI$ and $VE$ are generalisations of Pfenning's and Davies's definition of validity cast into rules of a system of natural deduction. Next we generalise the $I$- and $E$-rules for necessity. $\Box$ is treated as a multi-grade constant which has one formula to its right and 0 to finite $n$ formulas on its left. I abbreviate $B_1...B_n$ by $\Gamma$ and write $\Gamma\vdash A$ instead of $B_1...B_n\vdash A$ and $[\Gamma]^{\overline{i}}$ instead of $[B_1]^{i_1}...[B_n]^{i_n}$. The rules for $\Box$ are:

\begin{center}
\bottomAlignProof
\AxiomC{$[\Gamma]^{\overline{i}}$}
\noLine 
\UnaryInfC{$\Pi$}
\noLine
\UnaryInfC{$A$}
\RightLabel{$_{\overline{i}}$}
\LeftLabel{$\Box I:$}
\UnaryInfC{$\Gamma\Box A$}
\DisplayProof\qquad 	
\bottomAlignProof
\AxiomC{$\Sigma$}
\noLine
\UnaryInfC{$\Gamma\Box A$} 
\AxiomC{$[\Gamma \vdash A]^i$}
\noLine
\UnaryInfC{$\Xi$}
\noLine
\UnaryInfC{$C$}
\LeftLabel{$\Box E:$}
\RightLabel{$_i$}
\BinaryInfC{$C$}
\DisplayProof 
\end{center}

\noindent where in $\Box I$, all open assumptions of level 0 of $\Pi$ are in $\Gamma$ (any other open assumptions are of level 1 and have the form $\Delta \vdash B$). Vacuous discharge is allowed. In $\Box E$, $C$ is a 0-level formula. I propose to read $\Box$ as relative necessity. It expresses that there is a necessary connection between the formulas in $\Gamma$ and $A$, or necessarily, $A$ given $\Gamma$.\footnote{For a few more thoughts on this modal notion, see \citep{kurbisPTSNM}. It should be noted that on this reading, $\top$ is necessary relative to everything, while everything is necessary relative to $\bot$: the notion of relative necessity proposed here is not a relevant relative necessity.} When $\Gamma$ is empty, we get the usual unary necessity operator: it behaves as in intuitionist \textbf{S4}. 

Maximal formulas of the form $\Gamma\vdash A$ are removed by the following reduction step:

\begin{center} 
\AxiomC{$[B_1]^{i_1}...[B_n]^{i_n}$}
\noLine  
\UnaryInfC{$\Pi$}
\noLine
\UnaryInfC{$A$} 
\RightLabel{$_{i_1...i_n}$}
\UnaryInfC{$B_1...B_n \vdash A$}
\AxiomC{$\Xi_1$}
\noLine
\UnaryInfC{$B_1$}
\AxiomC{$...$}
\AxiomC{$\Xi_n$}
\noLine
\UnaryInfC{$B_n$}
\QuaternaryInfC{$A$}
\noLine
\UnaryInfC{$\Sigma$} 
\DisplayProof\quad$\leadsto$\quad
\def\defaultHypSeparation{\hskip .001in}
\AxiomC{$\Xi_1$}
\noLine
\UnaryInfC{$[B_1]$}
\AxiomC{$...$}
\AxiomC{$\Xi_n$}
\noLine
\UnaryInfC{$[B_n]$}
\noLine
\TrinaryInfC{$\Pi$}
\noLine
\UnaryInfC{$A$}
\noLine
\UnaryInfC{$\Sigma$} 
\DisplayProof
\end{center}
  
\noindent The restrictions on $VI$ and $\Box I$ require that all open formulas or all open 0-level formulas are discharged above their premises, and hence there can be no application of these rules in $\Pi$ below $B_1...B_n$, except where an assumption class $[B]^i$ is empty. So the transformation cannot lead to any violations of restrictions on rules in $\Pi$. Any applications of those rules also remain correct in $\Sigma$, as the reduction procedure does not introduce new open assumptions into the deduction. For essentially the same reason, Prawitz's reduction procedures for maximal formulas and segments continue to work for the constants \textbf{HM} shares with \textbf{I}. 

Removing maximal formulas $\Gamma\Box A$ is slightly more original: 

\begin{center}
\AxiomC{$[\Gamma]^{\overline{i}}$}
\noLine 
\UnaryInfC{$\Pi$}
\noLine
\UnaryInfC{$A$}
\RightLabel{$_i$}
\UnaryInfC{$\Gamma\Box A$}
\AxiomC{$[\Gamma \vdash A]^j$}
\AxiomC{$\overline{\Sigma}$}
\noLine
\UnaryInfC{$\Gamma$}
\BinaryInfC{$A$}
\noLine
\UnaryInfC{$\Xi$}
\noLine
\UnaryInfC{$C$}
\RightLabel{$_j$}
\BinaryInfC{$C$}
\noLine
\UnaryInfC{$\Theta$} 
\DisplayProof \qquad $\leadsto$ \qquad
\AxiomC{$\overline{\Sigma}$}
\noLine
\UnaryInfC{$[\Gamma]$}
\noLine 
\UnaryInfC{$\Pi$}
\noLine
\UnaryInfC{$A$}
\noLine
\UnaryInfC{$\Xi$}
\noLine
\UnaryInfC{$C$}
\noLine
\UnaryInfC{$\Theta$} 
\DisplayProof
\end{center}

\noindent $\overline{\Sigma}$ are the deductions of the formulas in $\Gamma$. A maximal formula of type $\Gamma\Box A$ can only occur in the context on the left, unless $\Gamma\vdash A$ is discharged vacuously by $\Box E$, in which case its removal is trivial. The only thing one can do with $\Gamma\vdash A$ is to apply $VE$ to it. Due to the restriction on $C$ in $\Box E$ and the formation rules for the language of \textbf{HM}, such a formula cannot be assumed and immediately discharged by a rule. Due to the restrictions on $VI$ and $\Box I$, there can be no applications of these rules below the $\Gamma$s in $\Pi$ (unless in the case of vacuous discharge, which is trivial): hence concluding the $\Gamma$s with the deductions in $\overline{\Sigma}$ cannot lead to violations of rules in $\Pi$. Due to the restrictions on $VI$, there can be no application of that rule below $A$ in $\Xi$, as there is at least the open assumption $\Gamma\vdash A$ that prevents such an application. If there is an application of $\Box I$ in $\Xi$, then all open assumptions of the deductions in $\overline{\Sigma}$ are of the form $\Delta \vdash B$, and hence they remain correct after the transformation. For similar reasons, applications of these rules in $\Theta$ also remain correct. 

All reduction steps reduce the complexity of the deduction: a maximal segment is shortened, a maximal formula of higher degree than those that may be introduced by the reduction procedure removed. A standard induction over the complexity of deductions establishes the normalisation theorem for \textbf{HM}.

\section{Conclusion} 
\textbf{HM} is a natural system of modal logic with higher order rules. It fulfils necessary conditions for a proof-theoretic account of the meaning of $\Box$. Deductions normalise. Its rules are harmonious and satisfy the molecularity principle. The meaning of $\Box$ is given in terms of the meaning of $\vdash$, the meaning of which is given in terms of inferences in \textbf{I}. 

\textbf{HM} generalises \textbf{JM} in introducing a more general notion of validity and allowing validities to occur as conclusions of rules. But it remains close to \textbf{JM}, in that the restrictions on $VI$ and the rules for $\Box$ are directly lifted from \textbf{JM}. A natural question is how the restrictions on $VI$ could be loosened to allow further ways of deriving $\Gamma\vdash A$. The restriction on $VI$ blocks a derivation of a version of cut: If (1) $\Gamma\vdash A$ and (2) $\Delta, A\vdash C$, then (3) $\Gamma, \Delta\vdash C$. It is possible to conclude $A$ from (1) by assuming all formulas in $\Gamma$, and then to conclude $C$ from (2) by assuming all formulas in $\Delta$, but the restriction on $VI$ prevents the conclusion of (3), as besides the 0-level formulas in $\Gamma$ and $\Delta$, the conclusion $C$ depends on the undischarged first level formulas $\Gamma\vdash A$ and $\Delta, A\vdash C$. 

Do\v sen proposes systems of higher order sequents for intuitionist and classical \textbf{S4} and \textbf{S5} (see \citep{dosensequentmodal}, \citep{dosenintmodal}), in which, he explains, $\Box A$ means `$A$ is assumed as a theorem'. This sounds similar to Pfenning's and Davies's account of modality. Do\v sen's system implements a stricter distinction of levels of formulas and rules than \textbf{HM}. To the left and right of Do\v sen's turnstile of level 2, there must be formulas of level 1, not of level 1 or 0. Thus transposed into a system of natural deduction, Do\v sen's rules for $\Box$, which are of level 2, would require premises and conclusions of level 1. These rules are derivable using present the rules if $VI$ may also be applied when all assumptions on which its premise depends are of level 1, i.e. of form $\Delta \vdash C$. Furthermore, with the restriction on $VI$ so loosened that amongst the assumptions on which its premise depends there may be formulas of level 1, the version of cut mentioned in the previous paragraph becomes derivable. Modifying \textbf{HM} is an avenue for further research. 

\bigskip 

\setlength{\bibsep}{0pt}
\bibliographystyle{chicago}
\bibliography{aiml20}

\end{document}